\documentclass{elsart}

\usepackage{amsmath}
\usepackage{amsfonts}            
\usepackage{amssymb}

\usepackage{mathrsfs}

\usepackage{eufrak}

\usepackage{graphicx}% Include figure files
\usepackage{epstopdf}% Include figure files

\usepackage{bm}% bold math

\begin{document}

\begin{frontmatter}

\title{The effect of a nonresonant radiative field on low-energy rotationally inelastic \\ $\text{Na}^{+}~+~\text{N}_2$ collisions}

\author{Mikhail Lemeshko}, \author{Bretislav Friedrich}
%\ead{lemeshko@fhi-berlin.mpg.de}

%\ead{brich@fhi-berlin.mpg.de}

\address{Fritz-Haber-Institut der Max-Planck-Gesellschaft\\ Faradayweg 4-6, D-14195 Berlin, Germany}

\date{\today}% It is always \today, today,
             %  but any date may be explicitly specified
\begin{abstract}

We examine the effects of a linearly polarized nonresonant radiative field on the dynamics of rotationally inelastic $\text{Na}^{+}~+~\text{N}_2$ collisions at eV collision energies. Our treatment is based on the Fraunhofer model of matter wave scattering and its recent extension to collisions in electric fields [arXiv:0804.3318v1]. The nonresonant radiative field changes the effective shape of the target molecule by aligning it in the space-fixed frame. This markedly alters the differential and integral scattering cross sections. As the cross sections can be evaluated for a polarization of the radiative field collinear or perpendicular to the relative velocity vector, the model also offers predictions about steric asymmetry of the collisions. 
\end{abstract}

\end{frontmatter}

%\maketitle

\section{Introduction}

Reactions between ions and simple molecules have been invoked in the chemistry of
comets~\cite{comets}, dense interstellar clouds~\cite{ionmoleculeinterstellar}, as well as in
the atmospheric chemistry of planet-like objects, such as Io~\cite{Io} and Titan~\cite{Titan}. Since the late 1990s, the Na$^+~+~$X collisions (with X an atmospheric ligand) have been recognized to be responsible for the formation of sporadic sodium layers in Earth's upper
mesosphere~\cite{SodLay}.

The species that take part in the reactions in the upper
atmosphere and in interstellar space are exposed to electromagnetic radiation of varying intensity. When polarized, this
radiation may create directional molecular states in which the spatial distribution of the molecular axis is itself
polarized. This axis polarization arises from the nonresonant interaction of the radiation with the anisotropic molecular polarizability
\cite{FriHer}. Here we examine how such polarization may affect the differential and integral
cross sections of the ion-molecule collisions.

In particular, we investigate the effect of an intense nonresonant radiative field on the
rotationally inelastic collisions of Na$^{+}$ ions with N$_2$ molecules at eV collision
energies. This collision system is of paramount importance  in generating sporadic sodium in the mesosphere~\cite{SodLay}.

In our investigation, we make use of a recently developed quantum model of collisions in fields \cite{LemFri}, which we here adapt for the case of an induced-dipole interaction of a nonresonant radiative field with molecular polarizability \cite{FriHer}. We limit our considerations to the scattering of $^1\Sigma$ molecules by ground-state atomic ions. The model is based on Fraunhofer scattering of matter waves \cite{Drozdov}, \cite{Blair}, \cite{Faubel}, and is analytic in both its field-free and field-dependent variant. 

In Section~\ref{sec:FraunApprox}, we briefly describe the Fraunhofer model of matter-wave scattering and its extension to the case of scattering in nonresonant radiative fields. In Section ~\ref{sec:Na-N2}, we apply the model to the  $\text{Na}^{+}~+~\text{N}_2$ rotationally inelastic collisions, and evaluate their differential and integral cross sections and the steric asymmetry as a function of the intensity of the radiative field. The main conclusions of this work are summarized in Section~\ref{sec:conclusions}.

\section{The Fraunhofer model of matter-wave scattering}
\label{sec:FraunApprox}

\subsection{Field-free scattering}
\label{sec:fieldfree}

\begin{figure*}[htbp]
\centering\includegraphics[trim=180 220 250 75, clip,width=6cm]{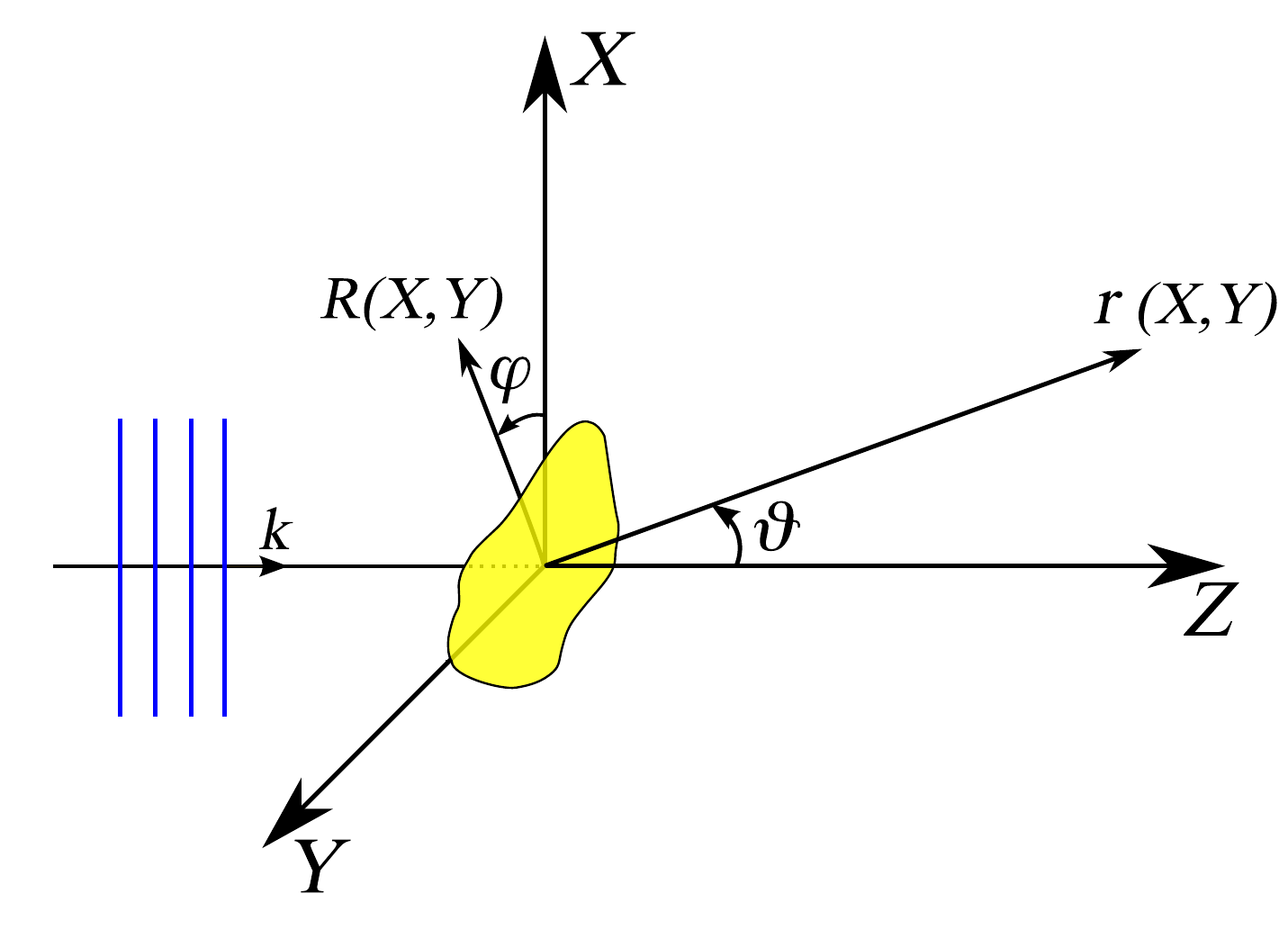}
\caption{Schematic of Fraunhofer diffraction by an impenetrable, sharp-edged obstacle as observed at a point of radius vector $\textbf{r}(X,Z)$ from the obstacle. Relevant is the shape of the obstacle in the $XY$ plane, perpendicular to the initial wave vector, $\mathbf{k}$, itself directed along the $Z$-axis of the space-fixed system $XYZ$. The angle $\varphi$ is the polar angle of the radius vector $\textbf{R}$ which traces the shape of the obstacle in the $X,Y$ plane and $\vartheta$ is the scattering angle. See text.}\label{fig:fraunhofer}
\end{figure*}

Inherent to the Fraunhofer model of matter-wave scattering is the energy sudden approximation and the assumption of an impenetrable, sharp-edged scatterer \cite{LemFri}. As a result, the Fraunhofer amplitude for scattering into an angle $\vartheta$ from an initial, $\vert \mathfrak{i} \rangle$, to a final,  $\vert \mathfrak{f} \rangle$, state is given by
\begin{equation}
	\label{InelAmplSudden}
	f_{\mathfrak{i} \to \mathfrak{f}} (\vartheta) = \langle   \mathfrak{f} \vert{f}(\vartheta) \vert \mathfrak{i} \rangle
\end{equation}
with 
\begin{equation}
	\label{FraunAmpl}
	f {(\vartheta)} \approx \int e^{-i k R \vartheta \cos \varphi} d R
\end{equation}
the amplitude for Fraunhofer diffraction as observed at a point of radiusvector $\textbf{r}$ from the scatterer, see Fig.~\ref{fig:fraunhofer}. Here $\varphi$ is the polar angle of the radius vector $\textbf{R}$ which traces the shape of the scatterer, $R\equiv|\mathbf R|$, and $k\equiv|\mathbf k|$ with $\mathbf k$ the initial wave vector. Relevant is the shape of the obstacle in the space-fixed $XY$ plane, perpendicular to $\mathbf{k}$, itself directed along the space-fixed $Z$-axis, cf. Fig.~\ref{fig:fraunhofer}. 

We note that the notion of a sharp-edged scatterer comes close to the rigid-shell approximation, widely used in classical~\cite{Beck79},~\cite{IchimuraNakamura},~\cite{Marks_ellips}, quantum~\cite{Bosanac}, and quasi-quantum~\cite{Stolte} treatments of field-free molecular collisions, where the collision energy by far exceeds the depth of any potential energy well.

In optics, Fraunhofer (i.e., far-field) diffraction~\cite{BornWolf} occurs when the Fresnel number is small,
\begin{equation}
	\label{FresnelNumber}
	\mathscr{F} \equiv \frac{a^2}{r \lambda} \ll 1
\end{equation}
Here $a$ is the dimension of the obstacle, $r\equiv|\textbf{r}|$ is the distance from the obstacle to the observer, and $\lambda$ is the wavelength, cf. Fig.~\ref{fig:fraunhofer}. Condition~(\ref{FresnelNumber}) is well satisfied for nuclear scattering at MeV collision energies as well as for molecular collisions at thermal and hyperthermal energies. In the latter case, inequality~(\ref{FresnelNumber}) is fulfilled due to the compensation of the larger molecular size $a$ by a larger de~Broglie wavelength $\lambda$ pertaining to thermal molecular velocities.

For nearly-circular targets, with a boundary $R (\varphi) = R_0 +\delta(\varphi)$ in the $XY$ plane, the Fraunhofer integral of Eq.~(\ref{FraunAmpl}) can be evaluated and expanded in a power series in the deformation $\delta(\varphi)$,
\begin{equation}
	\label{AmplitudeExpansion}
	f_{}  {(\vartheta)}  = f_0 (\vartheta) + f_1 (\vartheta,\delta) + f_2(\vartheta,\delta^2)+\cdots
\end{equation}
with $f_0(\vartheta)$ the amplitude for scattering by a disk of radius $R_0$
\begin{equation}
	\label{AmplSphere}
	f_0 (\vartheta) = i (k R_0^2) \frac{J_1 (k R_0 \vartheta)}{(k R_0 \vartheta)}
\end{equation}
and $f_1$ the lowest-order anisotropic amplitude,
\begin{equation}
	\label{AmplFirstOrder}
	f_1(\vartheta) = \frac{i k}{2 \pi} \int_{0}^{2 \pi} \delta(\varphi) e^{- i (k R_0 \vartheta) \cos \varphi} d\varphi
\end{equation}
where $J_1$ is a Bessel function of the first kind. Both Eqs.~(\ref{AmplSphere}) and~(\ref {AmplFirstOrder}) are applicable at small values of $\vartheta \lesssim 30^{\circ}$, i.e., within the validity of the approximation $\sin \vartheta \approx \vartheta$. 

The scatterer's shape in the space fixed frame, see Fig.~\ref{fig:fraunhofer}, is given by
\begin{equation}
	\label{RhoExpSpaceFixed}
	R (\alpha, \beta, \gamma ; \theta, \varphi) = \sum_{\kappa \nu \rho} \Xi_{\kappa \nu} \mathscr{D}_{\rho \nu}^{\kappa} (\alpha \beta \gamma) Y_{\kappa \rho} (\theta, \varphi)
\end{equation}
where $(\alpha,\beta,\gamma)$ are the Euler angles through which the body-fixed frame is rotated relative to the space-fixed frame, $(\theta, \varphi)$ are the polar and azimuthal angles in the space-fixed frame, $\mathscr{D}_{\rho \nu}^{\kappa} (\alpha \beta \gamma)$ are the Wigner rotation matrices, and $\Xi_{\kappa \nu}$ are the Legendre moments describing the scatterer's shape in the body-fixed frame. Clearly, the term with  $\nu=0$ corresponds to a disk of radius $R_0$,
\begin{equation}
	\label{R0viab}
	R_0 \approx \frac{\Xi_{00}}{\sqrt{4\pi}}
\end{equation}
Since of relevance is the shape of the target in the $XY$ plane, we set $\theta=\frac{\pi}{2}$ in Eq.~(\ref{RhoExpSpaceFixed}). As a result,
\begin{equation}
\label{deltaphi}
	\delta(\varphi)=R (\alpha, \beta, \gamma ; \tfrac{\pi}{2}, \varphi)-R_0=R (\varphi) - R_0=\underset{\kappa \neq 0 }{ \sum_{\kappa \nu \rho}} \Xi_{\kappa \nu} \mathscr{D}_{\rho \nu}^{\kappa} (\alpha \beta \gamma) Y_{\kappa \rho} (\tfrac{\pi}{2}, \varphi)
\end{equation}
By combining Eqs.~(\ref {InelAmplSudden}), (\ref{AmplFirstOrder}), and (\ref{deltaphi}) we finally obtain
\begin{equation}
	\label{InelAmplExpress}
	f_{\mathfrak{i} \to \mathfrak{f}} (\vartheta) \approx \langle \mathfrak{f} \vert f_0 + f_1 \vert \mathfrak{i} \rangle = \langle \mathfrak{f} \vert f_1 \vert \mathfrak{i} \rangle =  \frac{i k R_0}{2 \pi}  \underset{\kappa+\rho~\textrm{even}}{\underset{\kappa \neq 0 } {\sum_{\kappa \nu \rho}}} \Xi_{\kappa \nu} \langle \mathfrak{f} \vert \mathscr{D}_{\rho \nu}^{\kappa} \vert \mathfrak{i} \rangle F_{\kappa \rho} J_{\vert \rho \vert} (k R_0 \vartheta)
\end{equation}
where
\begin{equation}
	\label{Flamnu}
	F_{\kappa \rho} = \left \{ 	\begin{array}{ccl}
		(-1)^{\rho} 2\pi \left( \frac{2\kappa+1}{4\pi} \right)^{\frac{1}{2}} (-i)^{\kappa} \frac{\sqrt{(\kappa+\rho)! (\kappa-\rho)! }}{(\kappa+\rho)!! (\kappa-\rho)!! }  &    &    \textrm{ for $\kappa+\rho$ ~even~and~ $\kappa \ge \rho$} 
		\\ \\						0     &    &     \textrm{ elsewhere} 
	\end{array}    \right .
\end{equation}
For negative values of $\rho$, the factor $(-i)^{\kappa}$ is to be replaced by $i^{\kappa}$.

%-------------------------------------------------------------------------------------------------------

\subsection{Scattering in a radiative field}
\label{sec:fielddependent}

When subject to an external electric field, the electronic distribution of any molecule becomes polarized to some extent. This interaction, governed by the molecular polarizability, results in an induced dipole moment. While for the experimentally feasible static fields such induced moments are very weak, sizable dipole moments can be induced by a radiative field. If the induced-dipole interaction is anisotropic and sufficiently strong, the molecular rotational states undergo \textit {hybridization} (coherent linear superposition) which aligns the molecular axis along the field vector~\cite{FriHer}. The strength of the interaction is characterized by a dimensionless parameter $\Delta \omega$
\begin{equation}
	\label{omegapar}
	\Delta \omega \equiv \frac{2\pi \Delta \alpha I}{Bc}=\frac{\Delta \alpha \varepsilon^2}{4B}
\end{equation}
with $\Delta \alpha = \alpha_{\parallel} - \alpha_{\perp}$ the polarizability anisotropy,  $\alpha_{\parallel,\perp}$ the polarizability components parallel and perpendicular to the molecular axis, $B$ the rotational constant of the molecule, $I$ the radiation intensity, and $\varepsilon$ the amplitude of the corresponding oscillating electric field. The induced-dipole interaction couples states of the free-rotor basis set with same $M$ but with $J$'s that differ by $0,\pm2$. Thus the resulting hybrid states take the form
\begin{align}
	\label{PendularStateGeneral1}
	\vert \tilde{J}, M; \Delta \omega \rangle &= \sum_{J=2n} a_{J M}^{\tilde{J}} (\Delta \omega) \vert J, M \rangle \hspace{0.9cm} \text{for $\tilde{J}$ even} \\
	\label{PendularStateGeneral2}
	\vert \tilde{J}, M; \Delta \omega \rangle &= \sum_{J=2n+1} a_{J M}^{\tilde{J}} (\Delta \omega) \vert J, M \rangle \hspace{0.5cm}  \text{for $\tilde{J}$ odd}
\end{align}
where $2n = m+ |M|$ and $2n+1 = m+ |M|$ with $m$ either $0,2,4\dots$ or $1,3,5 \dots$. The hybridization coefficients $a_{J M}^{\tilde{J}} (\Delta \omega)$ depend solely on the interaction parameter $\Delta \omega$. The symbol $\tilde{J}$ denotes the nominal value of $J$ that pertains to the field-free rotational state which adiabatically correlates with the hybrid state,
\begin{equation}
	\label{PendularToFreeRotor}
	\vert \tilde{J}, M, \Delta \omega \to 0 \rangle \to \vert  J, M \rangle
\end{equation}
Since the hybrid wavefunctions, Eqs.~(\ref{PendularStateGeneral1}) and~(\ref{PendularStateGeneral2}), comprise either even or odd $J$'s, the states have \textit{definite parity}, $(-1)^{\tilde{J}}$.

\begin{figure*}[htbp]
\centering\includegraphics[trim=20 420 20 100, clip,width=15.5cm]{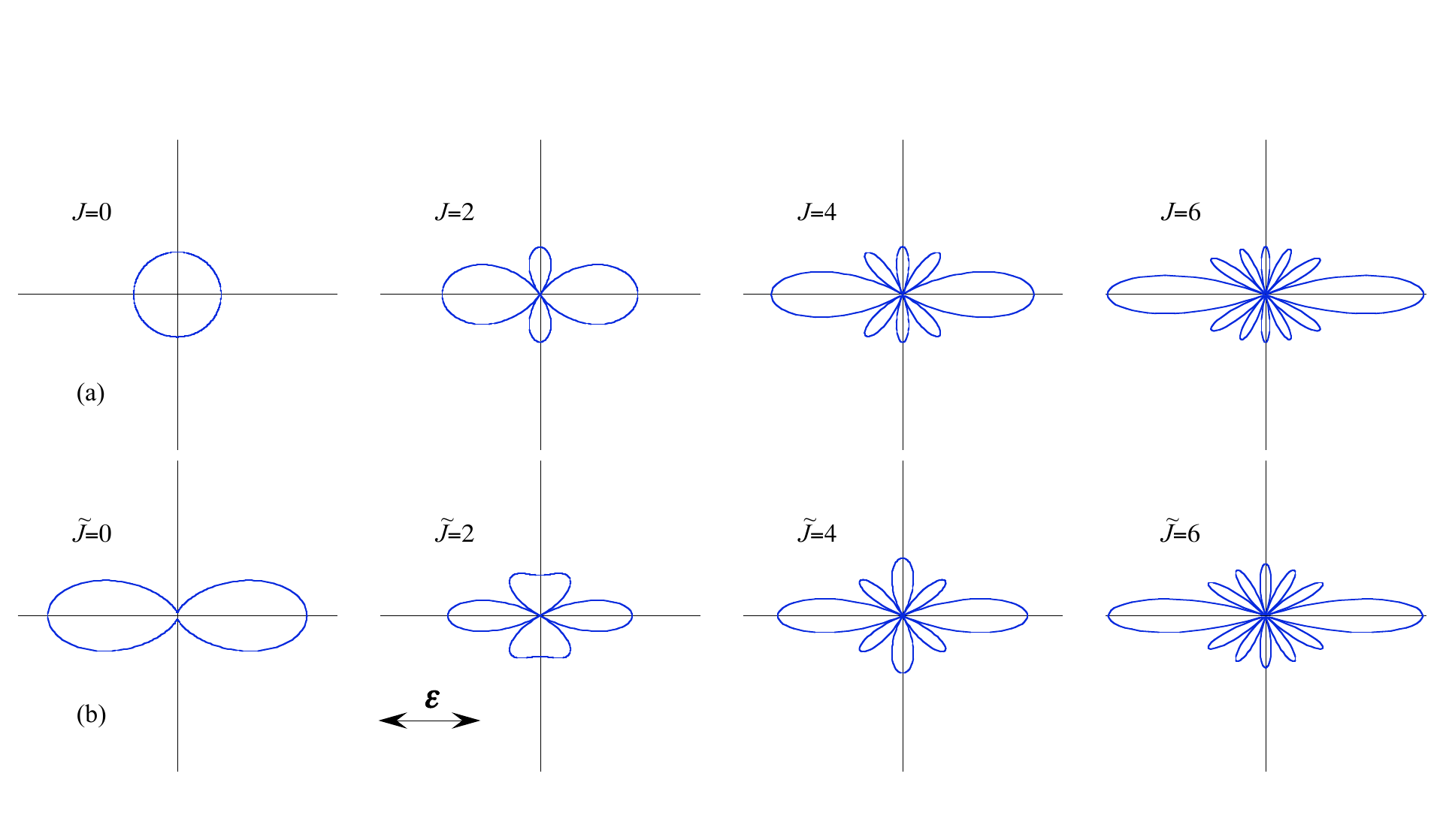}
\caption{A comparison of the moduli of the free rotor wavefunctions $\vert J, M=0 \rangle $, panel (a), with the moduli of the pendular wavefunctions $\vert \tilde{J}, M=0; \Delta \omega=25 \rangle $, panel (b). The polarization vector $\boldsymbol{\varepsilon}$ of the radiative field is also shown.}
\label{fig:pendstate}
\end{figure*}

Apart from possessing a particular energy level pattern, the $\vert \tilde{J}, M, \Delta \omega \rangle$ eigenstates are aligned along the electric field vector, $\boldsymbol{\varepsilon}$. The degree of alignment depends on the values of  $\tilde{J}$, $M$, and $\Delta \omega$. In such states, the molecular axis librates about the field direction like a pendulum, and so the hybrid states are referred to as \textit{pendular}. It is the directionality of the pendular states that enters the field-dependent Fraunhofer model and distinguishes it from the field-free model, which assumes an isotropic distribution of the molecular axes. The directional properties of pendular states are exemplified in Fig.~\ref{fig:pendstate}, which shows polar diagrams of both field-free and pendular wave functions at $\Delta \omega=25$.

The scattering process in the field consists of the following steps:  A molecule in a free-rotor state  $\vert J, M \rangle$ enters adiabatically the radiative field where it is transformed into a pendular state $\vert \tilde{J}, M, \Delta \omega \rangle$. This pendular state may be changed by the collision in the field into another pendular state, $\vert \tilde{J'}, M', \Delta \omega \rangle$. As the molecule leaves the field, the latter pendular state is adiabatically transformed into a free-rotor state $\vert J', M' \rangle$. Thus the net result is, in general,  a rotationally inelastic collision,  $\vert J, M \rangle \rightarrow \vert J', M' \rangle$.

In order to be able to apply Eq.~(\ref{InelAmplExpress}) to collisions in the radiative field,  we have to transform Eqs.~(\ref{PendularStateGeneral1}) and~(\ref{PendularStateGeneral2}) to the space-fixed frame $XYZ$. If the electric field vector is specified by the Euler angles $(\varphi_{\varepsilon},\theta_{\varepsilon},0)$ in the $XYZ$ frame, the initial and final pendular states take the form
\begin{align}
	\label{PendularStateASpFix}
	\vert \mathfrak{i} \rangle  \equiv \vert \tilde{J}, M; \Delta \omega \rangle & =  \sum_{J} a_{J M}^{\tilde{J}} (\Delta  \omega)\sum_{\xi} \mathscr{D}_{\xi M}^{J} (\varphi_{\varepsilon},\theta_{\varepsilon},0)  Y_{J \xi} (\theta,\varphi) \\
	\label{PendularStateBSpFix}
	\langle \mathfrak{f} \vert  \equiv \langle \tilde{J'}, M'; \Delta  \omega \vert & =  \sum_{J'} b_{J' M'}^{\tilde{J'} \ast} (\Delta \omega) \sum_{\xi'} \mathscr{D}_{\xi' M'}^{J' \ast} (\varphi_{\varepsilon},\theta_{\varepsilon},0)  Y_{J' \xi'}^{\ast} (\theta,\varphi)	
\end{align}
which is seen to depend solely on the angles $\theta$ and $\varphi$.

On substituting from Eqs.~(\ref{PendularStateASpFix}) and~(\ref{PendularStateBSpFix}) into Eq.~(\ref{InelAmplExpress}) and its integration, we obtain a general expression for the Fraunhofer scattering amplitude in the field,
\begin{multline}
	\label{ScatAmplArbField}	
	f_{\mathfrak{i} \to \mathfrak{f}}^{\omega} (\vartheta)= \frac{i k R_0}{2 \pi}  \underset{\kappa+\rho~\textrm{even}}{\underset{\kappa \neq 0 } {\sum_{\kappa,\rho}}} \mathscr{D}_{-\rho, \Delta M}^{\kappa \ast} (\varphi_{\varepsilon},\theta_{\varepsilon},0) \Xi_{\kappa 0} F_{\kappa \rho} J_{\vert \rho \vert} (k R_0 \vartheta)  \\
	\times \sum_{J J'} a_{J M}^{\tilde{J}} (\Delta \omega) b_{J' M'}^{\tilde{J'} \ast} (\Delta \omega) \sqrt{\frac{2J+1}{2J'+1}} C(J \kappa J' ; 0 0 0) C(J \kappa J' ; M \Delta M M')
\end{multline}
where $\Delta M \equiv M' - M$ and $C(J_1,J_2,J_3;M_1,M_2,M_3)$ are Clebsch-Gordan coeffients \cite{Zare}. Since the ion-linear molecule potential is axially symmetric, only the $\Xi_{\kappa 0}$ coefficients contribute to the scattering amplitude. 

Eq.~(\ref{ScatAmplArbField}) simplifies for special cases. If we limit our considerations to homonuclear diatomics, only the $\Xi_{\kappa 0}$ coefficients for even $\kappa$ contribute to the expansion, Eq.~(\ref{RhoExpSpaceFixed}), and, consequently, to the scattering amplitude, Eq.~(\ref{ScatAmplArbField}). Furthermore, if we fix the initial molecular state to the ground state, $\vert {J}, M \rangle \equiv \vert 0, 0 \rangle$, and restrict the polarization of the radiation in the space-fixed frame to a particular geometry, the problem simplifies as follows:

(i) For a polarization vector \textit{collinear} with the initial wave vector, $\boldsymbol{\varepsilon} \parallel \mathbf{k}$, we have $\theta_{\varepsilon}\rightarrow 0$, $\varphi_{\varepsilon} \rightarrow 0$. As as result, only the $\rho = - \Delta M'$ term yields a nonvanishing contribution and so	
\begin{multline}
	\label{Ampl0bParallel}
	f_{0,0 \to \tilde{J'}, M' }^{\omega, \parallel}  (\vartheta) = J_{\vert M' \vert} (k R_0 \vartheta) \frac{i k R_0}{2 \pi}  \underset{\kappa \neq 0} {\sum_{\kappa~\textrm{even}}}  \Xi_{\kappa 0} F_{\kappa M'} \\
	\times \sum_{J J'} a_{J 0}^{0} (\omega) b_{J' M'}^{\tilde{J'} \ast} (\omega) \sqrt{\frac{2J+1}{2J'+1}} C(J \kappa J' ; 0 0 0) C(J \kappa J' ; 0 M' M')
\end{multline}
We see that the angular dependence of the scattering amplitude for the parallel case is simple, given by a single Bessel function, $J_{\vert M' \vert}$.

(ii) If the polarization vector is \textit{perpendicular} to the initial wave vector,  $\boldsymbol{\varepsilon} \perp \mathbf{k}$, we have $\theta_{\varepsilon}\rightarrow \frac{\pi}{2}$,~$\varphi_{\varepsilon} \rightarrow 0$. Hence
\begin{multline}
	\label{Ampl0bPerp}
	f_{0,0 \to \tilde{J'}, M' }^{\omega, \perp} (\vartheta) = \frac{i k R_0}{2 \pi}  \underset{\kappa \neq 0 } {\sum_{\kappa,\rho~\textrm{even}}}  d_{-\rho, M'}^{\kappa}\left( \frac{\pi}{2} \right) \Xi_{\kappa 0} F_{\kappa \rho} J_{\vert \rho \vert} (k R_0 \vartheta) \\ 
	\times \sum_{J J'} a_{J 0}^{0} (\omega) b_{J' M'}^{\tilde{J'} \ast} (\omega)\sqrt{\frac{2J+1}{2J'+1}} C(J \kappa J' ; 0 0 0)  C(J \kappa J' ; 0 M' M')
\end{multline}
where $ d_{-\rho, M'}^{\kappa}$ are the real Wigner rotation matrices. Since the summation mixes different Bessel functions (for a range of $\rho$'s), the angular dependence of the scattering amplitude in the perpendicular case is more involved than in the parallel case.

The Clebsch-Gordan coefficient $C(J \kappa J' ; 0 0 0)$ in Eqs.~(\ref{Ampl0bParallel}) and~(\ref{Ampl0bPerp}) is nonzero only if $J+J'$ is even, since the summation includes only even-$\kappa$ terms. Moreover, given the definite parity of the pendular states, we see that only parity-conserving transitions are allowed, namely $J=0 \to J'=2,4,6,\dots$ for our choice of the initial state.

We can also see that, for either geometry, only the partial cross sections for the $J=0,M=0 \to J', M'$ collisions with $M'$ even contribute to the scattering. This is particularly clear in the $\boldsymbol{\varepsilon} \parallel \mathbf{k}$ case, where the $F_{\kappa M'}$ coefficients vanish for $M'$ odd. In the  $\boldsymbol{\varepsilon} \perp \mathbf{k}$ case, a summation over $\rho$ arises. Since for $\kappa$ even and $M'$ odd the real Wigner matrices obey the relation $d_{-\rho, M'}^{\kappa}\left( \frac{\pi}{2} \right) = - d_{\rho, M'}^{\kappa}\left( \frac{\pi}{2} \right)$, the sum over $\rho$ is zero and so are the partial cross sections with $M'$ odd.
\section{Rotationally inelastic collisions of Na$^{+}$ with N$_2$ in a radiative field}
\label{sec:Na-N2}

%As it was discussed in Sec.~\ref{sec:Fraun1Sigma}, the molecules with higher $\mu/B$ ratios exhibit stronger orientation in a given %electrostatic field $\boldsymbol{\varepsilon}$. Although such molecules as KCl, ICN or HCCCN are relatively heavy and have 
%large dipole moments, their employment in the experiments is unlikely (for a discussion see e.g. \cite{Friedrich99}).

Here we apply the model to the Na$^{+}$~+~N$_2 ( J=0 \to J')$ collisions. The polarization anisotropy $\Delta \alpha = 0.93$ \AA$^3$ and rotational constant $B = 1.9982$ cm$^{-1}$ make the N$_2$ molecule a suitable candidate for an experiment on laser-assisted ion-molecule collisions.

\begin{figure*}[htbp]
\centering\includegraphics[trim=50 200 50 200, clip,width=6.5cm]{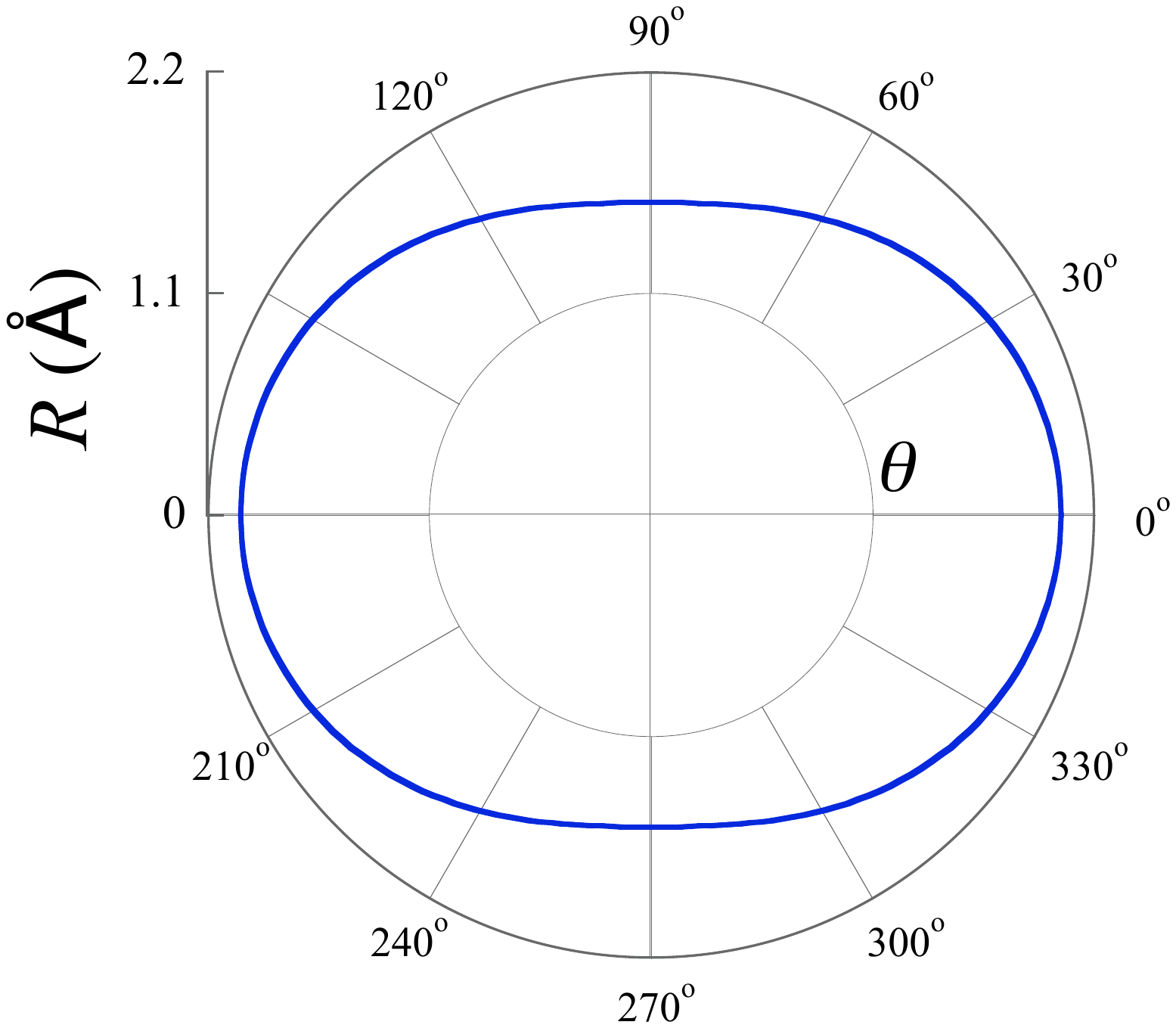}
\caption{Equipotential line $R(\theta)$ for the Na$^{+}$-- N$_2$ potential energy surface at a collision energy of 5~eV. The Legendre moments, Eq. (\ref{RhoExpSpaceFixed}), of the potential energy surface are listed in Table \ref{table:legendre_coefs}.}\label{fig:PEScut}
\end{figure*}

According to Ref.~\cite{Soldan}, the ground-state Na$^{+}$--N$_2$ potential energy surface has a global minimum  $-2712$~cm$^{-1}$ deep. The effect of this attractive well is negligible for low-energy collisions; we chose a collision energy of 5~eV, which corresponds to a wave number $k=173.8$~\r{A}$^{-1}$. The ``hard shell" of the potential energy surface at this collision energy is shown in Fig.~\ref{fig:PEScut}. We found it by a fit to Eq.~(\ref{RhoExpSpaceFixed}). The $\Xi_{\kappa 0}$ coefficients are listed in Table~\ref{table:legendre_coefs}. Due to the $D_{2v}$ symmetry of the potential energy surface,  only even-$\kappa$ terms arise. According to Eq.~(\ref{R0viab}), the $\Xi_{00}$ coefficient determines the hard-sphere radius $R_0$, responsible for elastic scattering. 

\subsection{Differential cross sections}
\label{diffcrossN2}

The \textit{field-free} state-to-state differential cross section, 
\begin{equation}
	\label{DiffCross00jmFF}
	\mathcal{I}_{0,0 \to J',M'}^{\text{f-f}}(\vartheta) =\vert f_{0,0 \to J',M'} (\vartheta) \vert^2 
	\end{equation}
see Eq. (\ref{InelAmplExpress}), is proportional to $\Xi^2_{J' 0}$, which means that the shape of the repulsive potential provides a direct information about the relative probabilities of the field-free transitions and \textit{vice versa}. Since for the Na$^{+}$--N$_2$ system the $\Xi_{2,0}$ coefficient dominates the anisotropic part of the potential, see Table~\ref{table:legendre_coefs}, the corresponding $J=0 \to J'=2$ collisions are expected to dominate the inelastic cross section. Because of the $D_{2v}$ symmetry, there are no parity-breaking $J=0 \to \text{odd } J'$ collisions in the  Na$^{+}$--N$_2 (J=0 \to J')$ system.

\begin{figure*}[htbp]
\centering
\includegraphics[trim=160 370 180 30, clip,width=8cm]{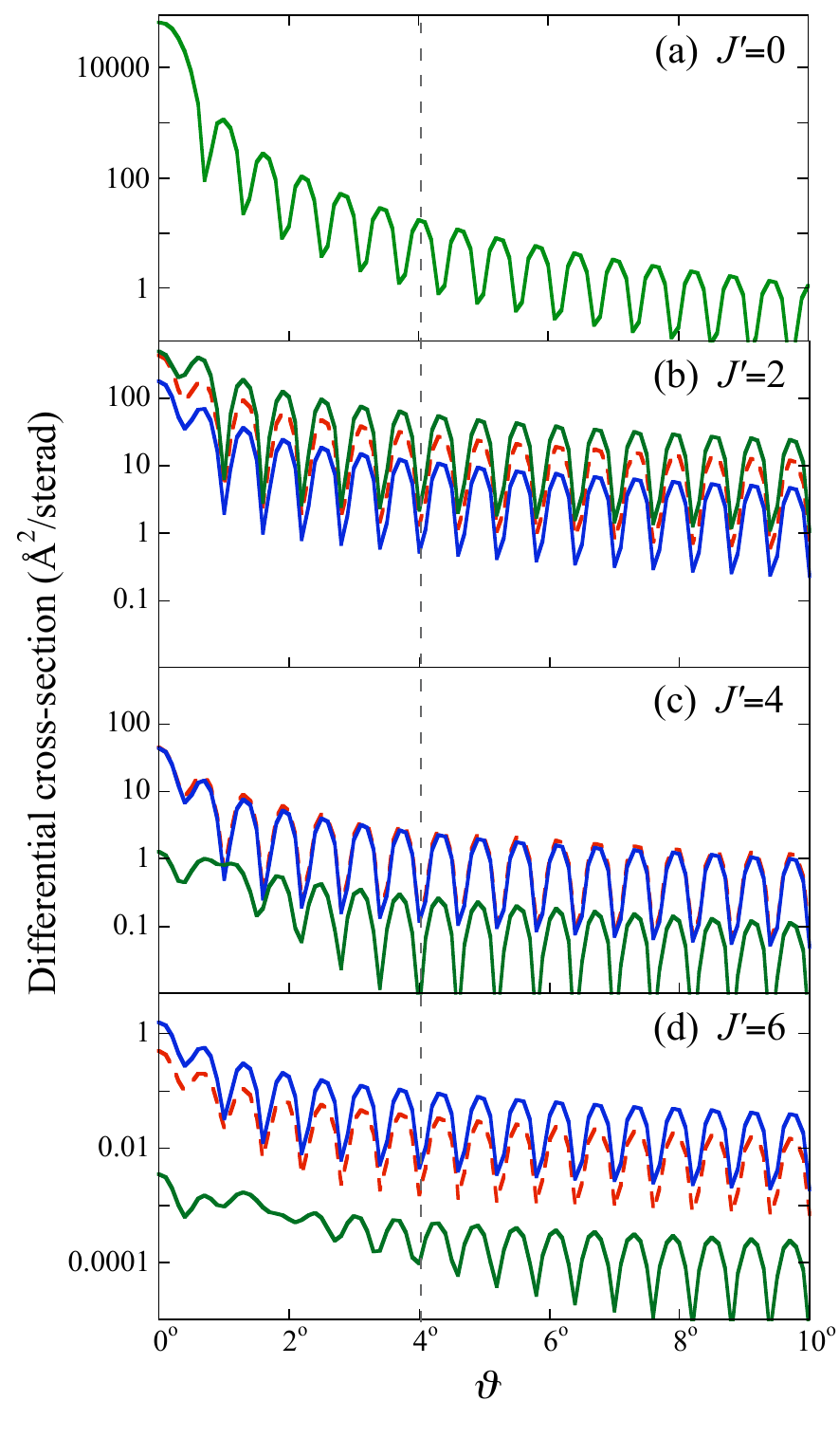}
\caption{Differential cross sections for the Na$^{+}$~+~N$_2$ $(J=0 \to J')$ collisions in a radiative field for $\Delta \omega = 10$ (red dashed line) and $\Delta \omega = 25$ (blue solid line), parallel to the initial wave vector. The field-free cross sections are shown by the green solid line.}
\label{fig:diff_parall}
\end{figure*}

\begin{figure*}[htbp]
\centering
\includegraphics[trim=160 370 180 30, clip,width=8cm]{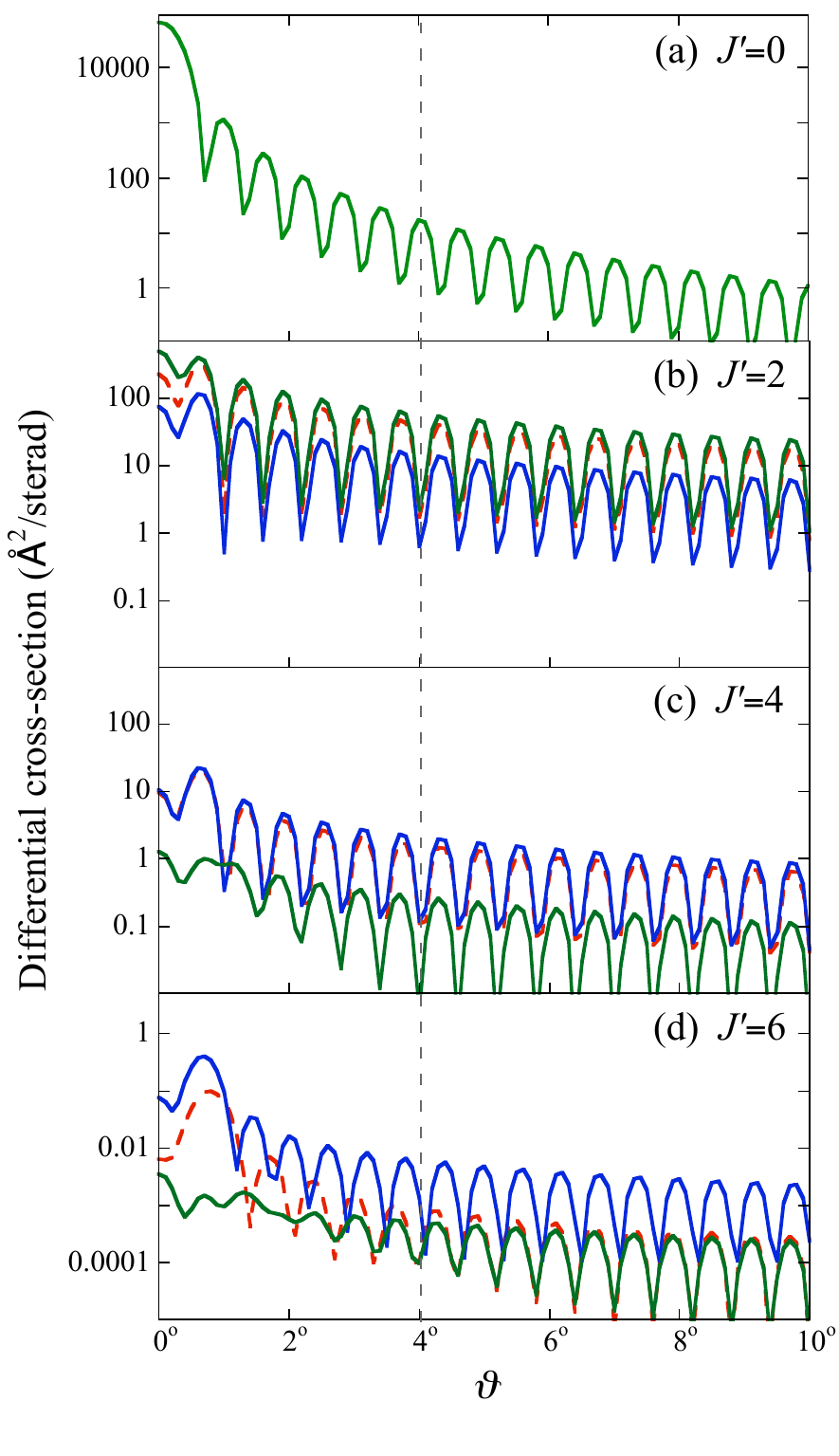}
\caption{Differential cross sections for the Na$^{+}$~+~N$_2$ $(J=0 \to J')$ collisions in a radiative field for $\Delta \omega = 10$ (red dashed line) and $\Delta \omega = 25$ (blue solid line), perpendicular to the initial wave vector. The field-free cross sections are shown by the green solid line.}
\label{fig:diff_perp}
\end{figure*}

After averaging over $M'$ and invoking the asymptotic properties of the Bessel functions \cite{Watson}, we obtain for the parity-conserving $J=0 \to \text{even } J'$ collisions
 \begin{equation}
	\label{AssympFiFrCrossJaver}
	\mathcal{I}_{0 \to J'}^{\text{f-f}}(\vartheta)  \sim \cos^2 \left(k R_0 \vartheta - \frac{\pi}{4}  \right)
\end{equation}
The elastic differential scattering cross section, cf. Eq.~(\ref{AmplSphere}), has a $\sin^2 \left(k R_0 \vartheta - \frac{\pi}{4} \right)$ asymptote, and so is seen to be shifted with respect to the differential cross sections for even-$J'$ transitions by a quarter of a wavelength. Known as the ``Blair phase rule," the shift is a conspicuous feature of Figs.~\ref{fig:diff_parall} and~\ref{fig:diff_perp}.
\\
\\The state-to-state differential cross sections for scattering in a radiative field parallel ($\boldsymbol{\varepsilon} \parallel \mathbf{k}$) and perpendicular ($\boldsymbol{\varepsilon} \perp \mathbf{k}$) to the initial wave vector are given by
\begin{equation}
	\label{DiffCrossFieldsJaver}
	\mathcal{I}_{0 \to J'}^{\omega,(\parallel,\perp)}(\vartheta)=\sum_{M'} \mathcal{I}_{0,0 \to J',M'}^{\omega,(\parallel,\perp)}(\vartheta)
\end{equation}
with
\begin{equation}
	\label{DiffCrossFieldsJM}
	\mathcal{I}_{0,0 \to J',M'}^{\omega,(\parallel,\perp)}(\vartheta)=\left \vert f_{0,0 \to \tilde{J'}, M' }^{\omega, (\parallel,\perp)} (\vartheta)  \right \vert^2
\end{equation}

The differential cross sections for the Na$^{+}$~+~N$_2$ collisions are presented in Figs.~\ref{fig:diff_parall} and~\ref{fig:diff_perp} for an interaction parameter $\Delta \omega = 10$ and $25$, corresponding to laser intensities of $2.15\times10^{12}$ W/cm$^2$ and $5.37\times10^{12}$ W/cm$^2$, respectively. The figures show that a radiative field on the order of 1$0^{12}$ W/cm$^2$ dramatically alters the magnitudes of the differential cross sections, but does not produce any ``phase shift" of the angular oscillations. Such a ``phase shift" is absent because only even Bessel functions, which have a $\cos^2 \left(k R_0 \vartheta - \frac{\pi}{4}  \right)$ asymptote, contribute to the scattering at any field strength, see Eqs.~(\ref{Ampl0bParallel})  and~(\ref{Ampl0bPerp}).

%--------------------------------------------------------------------------------------------------
\subsection{Integral cross sections}
\label{sec:IntCrossSec}

The angular range, $\vartheta \lesssim 30^{\circ}$, where the Fraunhofer approximation applies the best, comprises the largest impact-parameter collisions that contribute to the scattering the most, see Figs.~\ref{fig:diff_parall} and~\ref{fig:diff_perp}. Therefore, the integral cross section can be obtained to a good approximation by integrating the Fraunhofer differential cross section, Eq. (\ref{DiffCrossFieldsJaver}), over the solid angle $\sin\vartheta d \vartheta d \varphi$,
\begin{equation}
	\label{IntCrossSec}
	\sigma_{0 \to J'}^{\omega,(\parallel,\perp)} = \int_{0}^{2\pi} d\varphi \int_{0}^{\pi} \mathcal{I}_{0 \to J'}^{\omega,(\parallel,\perp)}(\vartheta) \sin\vartheta d \vartheta
\end{equation}
The integral cross-sections thus obtained for the field parallel and perpendicular to the initial wave vector are presented in Fig.~\ref{fig:integral_cross}.  One can see that the state-to-state cross section for the  $J=0 \to J'=2$ collisions steadily decreases with the interaction parameter $\Delta \omega$, whereas the other state-to-state cross sections show a non-monotonous dependence. These features can be explained by the field dependence of the overlap of the hybridization coefficients, $a_{J M}^{\tilde{J}} (\Delta \omega)$ and $b_{J' M'}^{\tilde{J'}}  (\Delta \omega)$, affecting the scattering amplitude, cf. Eqs.~(\ref{Ampl0bParallel}) and~(\ref{Ampl0bPerp}) and Ref. \cite{LemFri}.

\subsection{Steric asymmetry}
\begin{figure*}[htbp]
\includegraphics[trim=30 280 60 250, clip,width=15cm]{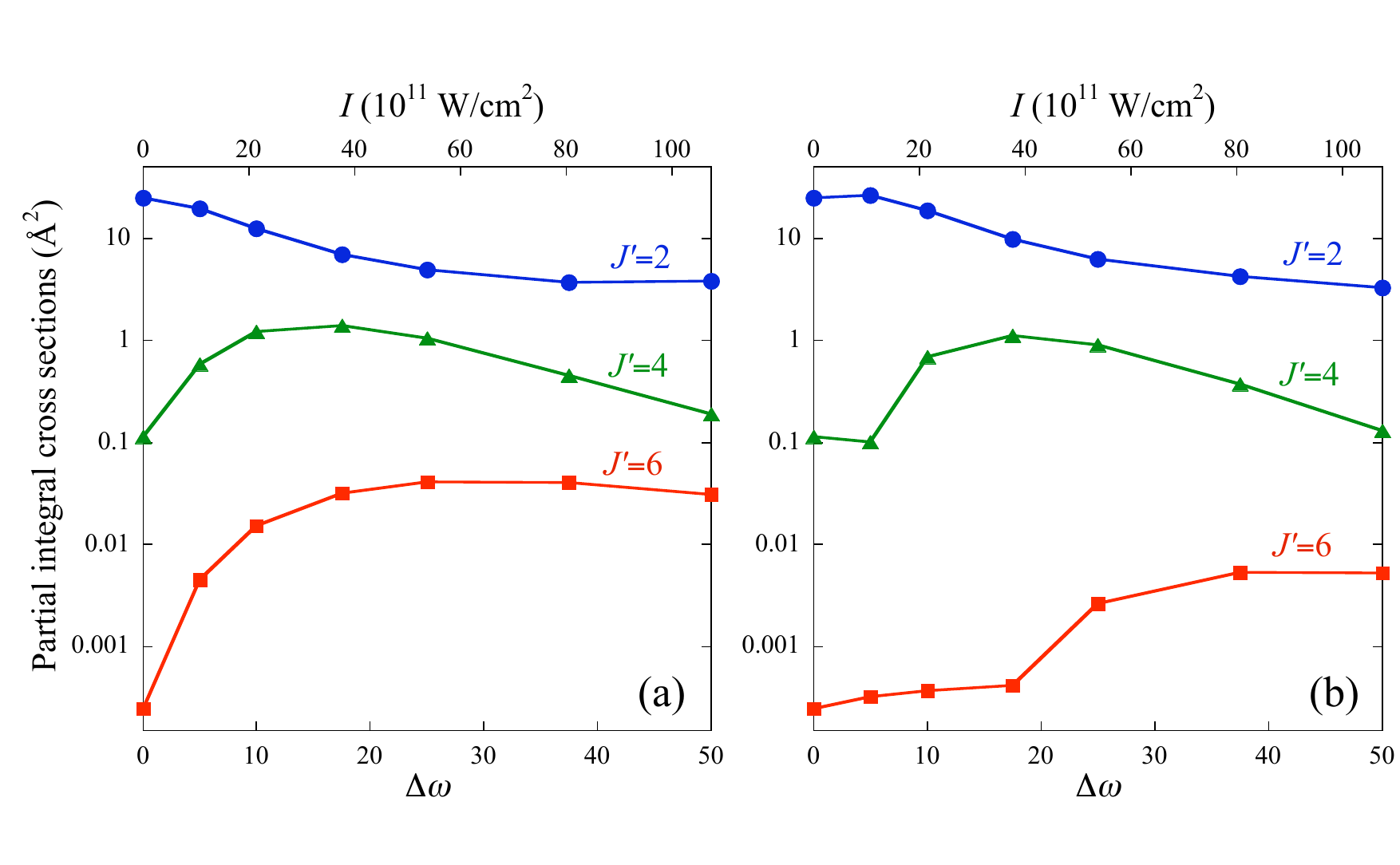}
\caption{Partial integral cross sections for Na$^{+}$~+~N$_2$ ($J=0 \to J'$) collisions in a radiative field parallel, panel~(a), and perpendicular, panel~(b), to the initial wave vector.}
\label{fig:integral_cross}
\end{figure*}

\begin{figure*}[htbp]
\centering\includegraphics[trim=45 170 40 140, clip,width=8cm]{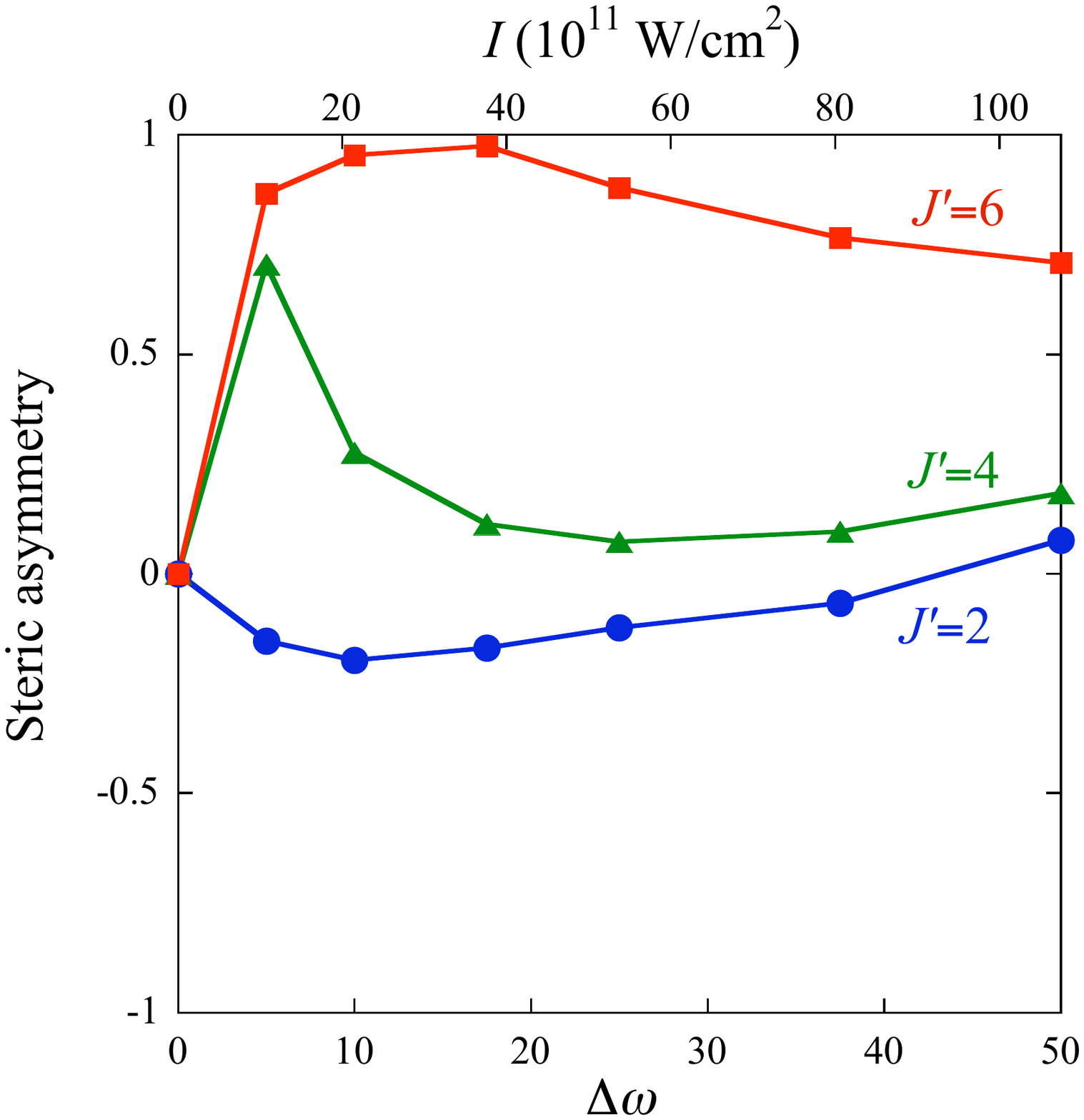}
\caption{Steric asymmetry, as defined by Eq.~(\ref{StericAsymmetry}), for Na$^{+}$~+~N$_2$ ($J=0 \to J'$) collisions.}\label{fig:asymmetry}
\end{figure*}

We define the steric asymmetry as
\begin{equation}
	\label{StericAsymmetry}
 	S_{\mathfrak{i} \to \mathfrak{f}} = \frac{\sigma_{\parallel}-\sigma_{\perp}}{\sigma_{\parallel}+\sigma_{\perp}},
\end{equation}
where the cross sections $\sigma_{\parallel, \perp}$ correspond, respectively, to $\boldsymbol{\varepsilon} \parallel \mathbf{k}$ and $\boldsymbol{\varepsilon} \perp \mathbf{k}$. The dependence of the steric asymmetry on the induced dipole interaction parameter $\Delta \omega$  is presented in Fig.~\ref {fig:asymmetry}. One can see that a particularly pronounced asymmetry obtains for the $J=0 \to J'=6$ channel. This can be traced to the field dependence of the corresponding integral cross sections, Fig.~\ref{fig:integral_cross}. 
Indeed, a conspicuous feature seen in Fig.~\ref{fig:integral_cross} is the significant dependence of the $J=0 \to J'=6$ channel on the collision geometry. The integral cross section for the $J=0 \to J'=6$ channel is always greater for the $\boldsymbol{\varepsilon} \perp \mathbf{k}$ geometry than it is for $\boldsymbol{\varepsilon} \parallel \mathbf{k}$ because of the non-vanishing $d^{\kappa}_{-\rho, M'}(\frac{\pi}{2})$ Wigner matrices, cf. Eqs.~(\ref{Ampl0bParallel}) and~(\ref{Ampl0bPerp}). The asymmetry for the $J=0 \to J'=2$ or $4$ channels is less pronounced, as only terms with $M'$ up to 2 or 4 are involved for  $\boldsymbol{\varepsilon} \perp \mathbf{k}$.

We note that within the Fraunhofer model, elastic collisions do not exhibit any steric asymmetry. This follows from the isotropy of the elastic scattering amplitude, Eq. (\ref{Ampl0bPerp}), which depends on the radius $R_0$ only: a sphere looks the same from any direction.

\section{Conclusions}
\label{sec:conclusions}

We made use of the Fraunhofer model of matter wave scattering to treat rotationally inelastic ion-molecule collisions in nonresonant radiative fields. In accordance with the energy sudden approximation, inherent to the Fraunhofer model, the interaction must be dominated by repulsion, which is typically well satisfied for ion-molecule collisions down to  collision energies on the oder of 1~eV. The Fraunhofer model is also inherently quantum and, therefore, capable of accounting for interference and other non-classical effects. 
The effect of the radiative field enters the model via the directional properties of the molecular states created by the field. Even a small alignment of the molecules was shown to cause a large alteration of the differential and integral cross sections. The strength of the analytic model lies in its ability to separate dynamical and geometrical effects and to qualitatively explain the resulting scattering features. These include the angular oscillations in the state-to-state differential cross sections or the rotational-state dependent oscillations in the integral cross sections as a function of the intensity of the radiative field. 

We hope that the model will inspire new experimental work based on a combination of ion traps with molecular beams \cite{gerlich}.

\section{Acknowledgements}

We dedicate this paper to Zdenek Herman, with gratitude for his friendship in life and science. We thank Gerard Meijer for his support an encouragement and Dieter Gerlich for discussions and suggestions.

%===========================================================================

%===========================================================================

\newpage

\begin{table}[h]
\centering
\caption{Hard shell  Legendre moments $\Xi_{\kappa 0}$, Eq. (\ref{RhoExpSpaceFixed}), for the $\text{Na}^{+}-\text{N}_2$ potential at a collision energy of $5$~eV. All odd moments are zero.}
\vspace{0.5cm}
\label{table:legendre_coefs}
\begin{tabular}{| c | c |}
\hline 
\hline 
 $\kappa$ &  $\Xi_{\kappa 0}$ (\r{A}) \\[3pt]
 \hline
 0 &  6.1221 \\
 2 &  0.5301  \\
 4 & -0.0359  \\
 6 & 0.0022   \\
 8 & 0.0002   \\
 \hline
 \hline 
\end{tabular}
\end{table}

%===========================================================================

\newpage

\newpage

\newpage

\newpage

\newpage

\newpage

\newpage

\end{document}